\documentclass[aps,pra,twocolumn,showpacs,superscriptaddress]{revtex4-1}
\usepackage{amsmath,amssymb}
\usepackage[pdftex]{graphicx}
\usepackage{booktabs}
\usepackage{color}
\newcommand \beq{\begin{eqnarray}}
\newcommand \eeq{\end{eqnarray}}

\begin{document}

\title{Analytical approach to a bosonic ladder subject to a magnetic field }
\author{Shun Uchino}
\affiliation{RIKEN Center for Emergent Matter Science, Wako,
Saitama 351-0198, Japan}
\date{\today}

\begin{abstract}
  We examine
  a bosonic two-leg ladder model subject to
  a magnetic flux, and especially 
  focus on a regime where the lower energy band has two minima.
  By using a low-energy field theory approach, we study several issues
  discussed in the system: the
  existence of local patterns in density and current, chiral-current reversal,
  and the effect of a nearest-neighbor interaction along the rung direction.
  In our formalism, the local patterns are interpreted as a result of
  breaking of discrete symmetry.
  The chiral-current reversal occurs through a competition between
  a current component determined at a commensurate vortex density causing an enlargement of
  the unit cell, and
  another component, which is proportional to the magnetic field doping
  from the corresponding commensurate flux. 
  The nearest-neighbor interaction along the rung direction available with
  the technique on a synthetic dimension is shown to
  favor a population-imbalance solution in an experimentally relevant regime.
\end{abstract}

\pacs{67.85.-d,05.30.Jp}

\maketitle

\section{Introduction}
Synthetic gauge fields in cold atoms provide a platform such that the
effects of magnetic and electric fields and
spin-orbit couplings are examined with atomic systems in which each atom is charge neutral
\cite{RevModPhys.83.1523,goldman2014light,zhai}.
Recently, there has been a great interest in the
effects of synthetic gauge fields
in quasi-one dimensional systems such as a ladder,
which displays properties of both one and higher dimensions \cite{RevModPhys.83.1405},
in cooperation with experimental realizations
on an optical superlattice  \cite{atala2014observation}
and on a synthetic dimension \cite{mancini,stuhl}.

In the presence of a magnetic flux,
vortices may penetrate into the system, which leads to an enlargement of
the unit cell and causes nontrivial phenomena.
The typical example is 
the Hofstadter butterfly originally considered
in the non-interacting two dimensional system \cite{PhysRevB.14.2239}.
The ladder geometry may be an ideal testing ground
to reveal roles of interactions in such a system since the powerful techniques
used in one dimension are  available \cite{giamarchi2003quantum}. 

The flux ladder has been analyzed in the context of
the condensed matter, which is mostly on fermions
\cite{PhysRevB.73.195114,PhysRevB.76.195105,PhysRevB.85.035104}.
Some theoretical studies on a bosonic ladder have been done
in terms of a Josephson-junction ladder \cite{PhysRevB.33.3125,PhysRevB.42.4797,nishiyama,PhysRevB.64.144515}.
An advantage in cold atoms is that one can start with the Bose-Hubbard Hamiltonian
without any ambiguity
\cite{PhysRevA.85.041602,PhysRevB.87.174501,PhysRevLett.111.150601,tokuno,PhysRevA.89.023619,PhysRevA.89.063617,PhysRevA.91.013629,PhysRevB.91.054520,PhysRevB.91.140406,PhysRevA.92.013625,PhysRevLett.115.190402,PhysRevA.92.053623,kolley,PhysRevB.92.060506,barbarino1,barbarino2}.
As the band structure  evolves as a function of a magnetic flux,
a weak-coupling regime in such a model is already nontrivial.
In particular, the bosons are sensitive to a shape of the band,
since they tend to occupy states around
the bottom of the lowest band. Therefore, one can expect a drastic change of
the low-energy properties due to the band evolution caused by
a flux.

The purpose of this paper is to diagnose
a weak-coupling regime of the bosonic flux ladder
in the light of a formalism developed in
Ref. \cite{PhysRevA.92.013625}.
By looking at a regime where
Bose-Einstein condensation is absent due to quantum fluctuations,
we discuss which properties predicted at the mean-field level
should be retained or changed.
It is shown that local patterns in density and current are
obtained as a consequence of breaking of discrete symmetry,
which is allowed in one dimensional systems.
We also discuss the chiral-current reversal recently obtained with
the density matrix renormalization group (DMRG) simulation
\cite{PhysRevLett.115.190402}. We demonstrate with our formalism that
such a chiral current reversal occurs through a competition
between a constant current component determined at a commensurate vortex
density and another current component that is proportional to
the magnetic field doping from the corresponding magnetic flux
and has the opposite sign to the former component.
Motivated by the recent experimental realization of the
synthetic dimension, we examine the role of an interchain
interaction, which turns out to encourage a population-imbalance
solution.

In Sec. II, we introduce the model discussed in this paper
and review  several analyses and properties in
a weak-coupling regime.
Section III explains our formalism and mentions a similarity to
fermion systems.
Results based on our formalism are shown in Sec. IV.
Section V is devoted to the summary.

\section{The model}
\label{sec:model}
The bosonic two-leg ladder system with a flux $\phi$
per plaquette is
defined as
\beq
&&H=-J\sum_{l=1}^{L}\sum_{p=1,2}(e^{-iA_{\parallel,p}}b^{\dagger}_{l+1,p}b_{l,p}+\text{H.c.})\nonumber\\
&&-K\sum_{l=1}^{L}(e^{-iA_{\perp}}b^{\dagger}_{l,1}b_{l,2}+\text{H.c.})
+\frac{U}{2}\sum_{l,p}n_{l,p}(n_{l,p}-1)\nonumber\\
&&+V\sum_{l}n_{l,1}n_{l,2},
\eeq
where
$p$ and $l$ respectively denote chain and site indices,
$b_{l,p}$ ($b^{\dagger}_{l,p}$)
represents the annihilation (creation)
operator of a boson at chain $p$ and site $l$,
and $n_{l,p}=b^{\dagger}_{l,p}b_{l,p}$.
The intrachain and interchain hoppings are denoted by $J$ and $K$, respectively,
and $U$ represents the on-site Hubbard interaction, which is
assumed to be positive in order to ensure the stability of the large system
size limit.
In the case of the synthetic dimension, 
an interchain interaction
 shows up due to
collisions between different internal states.
This coupling is denoted by $V$, and has a value similar 
to $U$ for the $^{87}$Rb case \cite{stuhl}
\footnote{To be precise, the on-site interaction $U$ can
  take a different value in each chain in the synthetic dimension
  case. Here, we neglect such a difference, for instance, by considering that
it is smaller than the temperature of the system.}.

In accord with the above Hamiltonian, 
the chiral and rung currents are defined as
\beq
&& j^c(l)=iJ(e^{-iA_{\parallel,1}}b^{\dagger}_{l+1,1}b_{l,1}-e^{-iA_{\parallel,2}}b^{\dagger}_{l+1,2}b_{l,2})+\text{H.c.}\\
&& j^{\perp}(l)=iKe^{-iA_{\perp}}b^{\dagger}_{l,1}b_{l,2}+\text{H.c.}
\eeq
which are
important to characterize different phases in the system.
In order to ensure that a magnetic flux is applied in the system, below,
we adopt the following gauge:
$A_{\parallel,1}=-A_{\parallel,2}=\frac{\phi}{2}$,
$A_{\perp}=0$.

\subsection{Phases in the absence of $V$}
The phase structure  in the present model was first
examined in studies of the Josephson-junction array
\cite{PhysRevB.33.3125,PhysRevB.64.144515}.
By assuming $J\gg K, U$ and using the bosonization method, 
the Meissner and vortex phases are obtained.
In the former phase realized in a low-field regime,
the rung current vanishes whilest
the chiral current increases with increasing  flux.
In the latter phase realized in a high-field regime, on the other hand,
the chiral current decreases with increasing flux
due to the penetration of vortices.
An implicit assumption used in Refs. \cite{PhysRevB.33.3125,PhysRevB.64.144515}
is that the mean density in each chain takes the same value, $n_1=n_2$.
However, it will turn out that this is the nontrivial assumption
in the presence of a flux.

Tokuno and Georges introduced another approach
where the two hoppings $J$ and $K$ are treated in a
non-perturbative manner \cite{tokuno}.
By diagonalizing the single particle Hamiltonian and
keeping degrees of freedom near the bottoms of the lower band,
the low-energy effective theory based on the bosonization method
has been discussed. 
In this analysis, the Meissner phase is obtained when
the lower band has a single minimum, while
the vortex phase is obtained when it has double minima.
It has also been shown that in the mean-field analysis at the atomic limit, 
the population-balance solution $n_1=n_2$ realized in the Meissner and
vortex phases exists.
In addition, it has  been mentioned  that
in a certain parameter regime, the Hessian matrix for the mean-field energy
becomes negative, implying that such a population-balance solution is unstable.

Wei and Mueller have explicitly shown with
the Gross-Pitaevskii (GP) approach
that a biased ladder phase, which has a population imbalance between the legs,
indeed exists \cite{PhysRevA.89.063617}.
In the biased ladder phase, the rung current vanishes, while
the chiral current decreases with increasing flux.
In fact, their analysis is similar to that in Ref. \cite{tokuno}
in the sense that the degrees of freedom around
the bottoms of the lower band are considered.
An essential difference is that they explicitly consider the following
form of a many-body ground state \cite{PhysRevA.89.063617}:
\beq
|\text{GS}\rangle=\frac{1}{\sqrt{N!}}(e^{i\bar{\theta}_+}\cos\gamma \beta^{\dagger}_{Q}
+e^{i\bar{\theta}_-}\sin\gamma \beta^{\dagger}_{-Q})^N|0\rangle,
\label{gp}
\eeq
where $\beta^{\dagger}_{\pm Q}$ is the creation operator of a boson
at the bottoms of the lower band $k=\pm Q$,
$\bar{\theta}_{\pm}$ is the phase factor,
and $\gamma$ describes a population rate between the minima.
Notice also that through a unitary transformation, 
$\beta_{k}$ is related to $b_{k,p}$, which is the Fourier transformation
of $b_{l,p}$.
Thus, it is natural to expect that the optimal ground state affects
the density profile in each chain.
Indeed, it has been shown that at a certain parameter regime,
$\gamma$ can be 0 or $\pi/2$, where
all the bosons occupy one of the minima and
spontaneous population imbalance between the legs occurs.

More recently, 
an extensive analysis based on the bosonization method has was
done in Ref. \cite{PhysRevA.92.013625}.
In this analysis, as in Refs. \cite{tokuno,PhysRevA.89.063617},
the effect of $J,K$ is incorporated in a non-perturbative manner
by diagonalizing the single particle Hamiltonian. Then,
the effect of the higher energy band is neglected by
considering the limit $J,K\gg U$.
A peculiar point in this analysis is that the bosonization 
based on the GP approach of Eq. \eqref{gp} is considered.

In general, the bosonization requires a mean-field solution
as a basis to consider quantum fluctuations.
Since the GP approach provides a correct basis
in the weak-coupling regime including the biased ladder phase,
the bosonization based on the GP approach can examine
quantum fluctuation effects in each phase with propriety.
For instance, 
the biased ladder phase is shown to be stable even in the presence
of quantum fluctuations \cite{PhysRevA.92.013625} while
Bose-Einstein condensation
predicted with the GP analysis
is replaced by the quasi-long range order described by
the Tomonaga-Luttinger liquid (TLL).
In terms of the symmetry, in the biased ladder phase,
the $Z_2$ symmetry breaking occurring at the mean-field level
is kept even at the quantum level,
while the U(1) symmetry broken at the mean-field level is recovered.
These properties in the biased ladder phase are also
confirmed by the recent DMRG simulation
\cite{PhysRevLett.115.190402}.

In the analysis in Ref. \cite{PhysRevA.92.013625},
the effect of the umklapp scattering between $k=\pm Q$
neglected in Ref. \cite{PhysRevA.89.063617}
was also incorporated.
When $Q$ takes commensurate values as
$Q=\pi p/q$ with coprime numbers $p,q$, such an effect is particularly important
and has an impact on the phase structure.
At the GP level, the ground-state energy starts to depend on the relative phase
$\bar{\theta}_+-\bar{\theta}_-$ due to the umklapp scattering
\cite{PhysRevB.87.174501,PhysRevA.92.013625}.
The relative phase acts as
an additional parameter and therefore
the phase boundary turns out to be shifted.
In addition, commensurability in $Q$ also affects the nature of the vortex
phase. The vortex phase realized at incommensurate $Q$, which is
called an incommensurate vortex or vortex fluid phase,
can be described by two-independent TLLs.
On the other hand, in the vortex phase realized at commensurate $Q$,
which is called a commensurate vortex or vortex lattice phase,
one of two TLLs becomes massive due to the umklapp scattering.
Thus, by changing a flux, which causes a shift of $Q$,
a transition between the commensurate vortex and incommensurate vortex phases
emerges. Then, such a transition in which
the GP analysis cannot access the nature, 
can be naturally explained by
the bosonization method in combination with the renormalization group
treatment
\cite{giamarchi2003quantum}.

In total, the analysis in Ref. \cite{PhysRevA.92.013625}
is useful to discuss low-energy properties 
of the flux ladder system with quantum fluctuations
in an analytic manner. 
In terms of this analysis, below,
we wish to discuss some issues of the bosonic flux ladder
by focusing on points not discussed in Ref. \cite{PhysRevA.92.013625}.

\section{Formalism}
As mentioned above, as far as the weak-coupling regime ($K,J\gg U,V$) is concerned,
bosons tend to populate the neighborhood of the bottom of the lower band,
and therefore, we may neglect the effect of the upper band.
The present model shows that
the lower band evolves as one with a single minimum to one with two minima
as the magnitude of the flux is increased.

In the case of a single minimum that corresponds to
the Meissner phase,
the elimination of the upper band leads to the relation
 $b_{l,p}\sim \frac{1}{\sqrt{2}}\beta_{l}$, where $\beta_l$
is the boson operator defined around the bottom of the lower band
at a site $l$
\cite{tokuno,PhysRevA.92.013625}.
Thus, the effective Hamiltonian can be represented as the single boson operator
$\beta_l$, and therefore
the problem is almost reduced to a single chain one without a flux.
The only difference from the case without a flux
is the emergence of a nonzero chiral current.
In addition, the bosonization method in the Meissner phase can be
straightforwardly performed as in the case of the single chain problem
\cite{tokuno,PhysRevA.92.013625}.
It is also shown that there is no local density or current pattern
in the Meissner phase.

In the case of the two minima where
the vortex and biased ladder phases can be obtained,
on the other hand,
more involved but rich situations appear.
There, two independent boson operators defined
around the two minima of the lower band play important roles in
the low-energy physics. 
To see this,
we note that by means of the long-wave length approximation,
we obtain the following relations in real space
\cite{tokuno,PhysRevA.92.013625}:
\beq
&&b_{l,1}= -e^{-iQl}\sin\frac{\xi_{Q}}{2}\beta_{l,-}
+e^{iQl}\cos\frac{\xi_{Q}}{2}\beta_{l,+},\label{eq:b1}\\
&&b_{l,2}= e^{-iQl}\cos\frac{\xi_{Q}}{2}\beta_{l,-}-
e^{iQl}\sin\frac{\xi_Q}{2}\beta_{l,+},
\label{eq:b2}
\eeq
where 
$\cos\frac{\xi_Q}{2}/\sin\frac{\xi_Q}{2}=\pm\sqrt{\frac{1}{2}\left(
  1\pm\frac{\sin Q\sin(\phi/2)}{\sqrt{(K/2J)^2+\sin^2Q\sin^2(\phi/2)}}\right)}$ are the weight factors by which
the single particle Hamiltonian can be diagonalized.
In the above, $\beta_{l,+(-)}$ is the bosonic operator at a site $l$ defined
around the right (left) bottom of the lower band.
In the formulation in \cite{tokuno,PhysRevA.92.013625},
the bosonization is performed not in $b_{l,p}$ but in $\beta_{l,\pm}$
as
\beq
\beta_{l,\pm}=\sqrt{n_{\pm}-\frac{\nabla\varphi_{\pm}}{\pi}}
\sum_{m\in Z}e^{2im[\pi n_{\pm}l-\varphi_{\pm}]}e^{i\theta_{\pm}},
\label{eq:bosonization}
\eeq
where $\varphi_{\pm}$ and $\theta_{\pm}$ are real fields describing the
density and phase fluctuations,
respectively\cite{giamarchi2003quantum}.
In our formalism,
the mean density $n_{\pm}$ can be determined by the GP solution.
In total, the fields $b_{l,p}$ are decomposed into
the slow fields $\varphi_{\pm}$ and $\theta_{\pm}$,
and rapidly oscillating factors $e^{\pm iQl}$.

The situation above is similar to
the bosonization of fermions in the sense that
in one dimensional systems,
the fermion operators can be decomposed into similar bosonic slow fields
describing the density and phase fluctuations, and
oscillating factors $e^{\pm ik_F l}$ with the Fermi momentum $k_F$
\cite{giamarchi2003quantum}.
Then, we naively obtain the TLL as the low-energy effective theory.
However, since  $k_F$ is related to the density,
commensurability of $k_F$ may lead to the instability of the TLL and the famous Mott insulator. 
In the bosonization language, this commensurability leads to
a cosine term on the density fluctuation field in the Hamiltonian,
and the low-energy effective theory
reduces to the famous sine-Gordon model. Then,
the Mott transition at half filling occurs
if the so-called TLL parameter is less than 1,
where the cosine term is relevant 
and the Mott gap is generated \cite{giamarchi2003quantum}.
When the situation in which the Mott insulator is realized is concerned,
by changing the chemical potential, which can be
represented as $\nabla\varphi$ in the bosonization,
the commensurate-incommensurate (C-IC) transition is allowed.
There, an infinitesimal change in chemical potential
does not give rise to this transition.
The C-IC transition
is allowed if the chemical potential added is of the order of
the Mott gap \cite{giamarchi2003quantum}.

Based on the above observations, we come back to Eqs.
\eqref{eq:b1},\eqref{eq:b2},\eqref{eq:bosonization}
to see fundamental rules in each parameter.
In the presence of the double minima in the lower band,
we can discuss the commensurability of $Q$. 
In this case, a cosine term on the (relative)
phase fluctuation field $\theta_{+}-\theta_-$ in the Hamiltonian emerges
through the terms containing $\beta^{\dagger}_{l,+}\beta_{l,-}$ and  $\beta^{\dagger}_{l,-}\beta_{l,+}$.
Since such a cosine term is relevant for a large TLL parameter
\cite{giamarchi2003quantum},
which is indeed
the case for the weakly-interacting bosons,
an instability of the TLL in this sector emerges.
Due to this mechanism, the commensurate vortex phase also called
vortex lattice phase is obtained.
As in the case of the Mott insulator, we can also discuss
the C-IC transition in the vortex phase.
There, the magnetic flux $\phi$ plays a similar role as the chemical potential
in the Mott insulator.
Namely, an infinitesimal change of $\phi$ does not invoke a change of
$Q$ (or vortex density) and the C-IC transition.
The change of $Q$ and C-IC transition is allowed if the change of $\phi$
exceeds the gap.

\section{Results}
Below, we discuss several properties in terms of our formalism
especially by focusing on the case of the band with the two minima
and points that are not discussed in Ref. \cite{PhysRevA.92.013625}.
In order to successfully apply the bosonization \eqref{eq:bosonization},
we first need to determine the mean density $n_{\pm}$. 
In our formalism, this can been done by the GP analysis
\cite{PhysRevA.89.063617,PhysRevA.92.013625}.
As mentioned above, there are two solutions:
population balance and imbalance ones.
Sections IV A, B, C discuss the $V=0$ case,
and the $V\ne0$ case is discussed in Sec. IV D.

\subsection{Population-imbalance solution}
The biased ladder phase belongs to the population-imbalance
solution where $n_+=n,$ $n_-=0$ or 
$n_+=0,$ $n_-=n$ is selected spontaneously.
While there are rapidly oscillating factors in Eqs.
\eqref{eq:b1} and \eqref{eq:b2},
these factors do not play any role in the density and current
expression since $\beta_{l,+}$ or $\beta_{l,-}$ is dropped in the biased
ladder phase.
Thus, the local patterns on the density and current are not expected.
As first pointed out in Ref. \cite{PhysRevA.89.063617},
the peculiarity of this phase is rather $n_{l,1}\ne n_{l,2}$
originating from $\sin\frac{\xi_{Q}}{2}\ne \cos\frac{\xi_Q}{2}$.

We also point out that at the transition point between
the biased ladder and vortex phases, an arbitrary imbalance
between $n_+$ and $n_-$ is allowed due to the emergent SU(2) symmetry
in the low-energy effective Hamiltonian
\cite{PhysRevA.92.013625}.
This point is interpreted as
the special case of the population-imbalance solution.

\subsection{Population-balance solution}
The mean density of the vortex phase is the same between
the two minima,
that is, $n_+=n_-=\frac{n}{2}$.
In this case, both of the rapidly oscillating terms, $e^{iQl}$
and $e^{-iQl}$ in Eqs. \eqref{eq:b1} and \eqref{eq:b2}
play an important role.

Let us now look at the local density profile following an analysis
in Ref. \cite{PhysRevA.89.063617}.
Within the GP analysis, one obtains
\beq
&&\langle n_{l,1} \rangle=\langle n_{l,2} \rangle=\nonumber\\
&&\frac{n}{2}-n\sin\frac{\xi_{Q}}{2}\cos\frac{\xi_{Q}}{2}\cos(2Ql-\bar{\theta}_++
\bar{\theta}_{-}),
\label{eq:density}
\eeq
where we note that in Ref. \cite{PhysRevA.89.063617},
$\bar{\theta}_+=\bar{\theta}_-=0$ is
implicitly
assumed. The above expression means that
except for $2Ql-\bar{\theta}_++\bar{\theta}_--=\frac{2m+1}{2}\pi$ ($m\in Z$),
the density in each leg is oscillating.
However, we point out that the above prediction requires considerable
attention and one must separately discuss the 
incommensurate and commensurate $Q$ cases.

In the incommensurate flux case, 
fixing $\bar{\theta}_+-\bar{\theta}_-$ at some constant is directly related to
the presence of an axial U(1) symmetry in the low-energy
Hamiltonian and its breaking at the mean-field level.
The axial U(1) symmetry represents that
the low-energy Hamiltonian
is invariant under the transformation
$\beta_{l,\pm}\to e^{\pm i\theta}\beta_{l,\pm}$ with a phase $\theta$
\cite{PhysRevA.92.013625}.
Then, the GP analysis predicts that the axial U(1) symmetry is
spontaneously broken.
However,
as far as the thermodynamic limit is concerned,  such a fixing does not occur
since the U(1) symmetry breaking in 1D system is not allowed
\footnote{The symmetry breaking may occur if one considers
  the effects of couplings to higher dimensions, finite systems, and
the trap, which may be relevant to the experiment.}.
To see this directly, let us examine quantum fluctuation effects
by means of the bosonization.
In such a prescription, 
the oscillating term in Eq. \eqref{eq:density} is bosonized as
$\cos(2Ql-\sqrt{2}\theta_a)$, where
$\theta_a=\frac{1}{\sqrt{2}}(\theta_+-\theta_-)$ is
not the constant but the field operator.
Since the low-energy effective Hamiltonian on $\theta_a$
is shown to be the TLL one \cite{PhysRevA.92.013625}, 
we found that the oscillation in local density predicted by
the mean-field theory disappears by
quantum fluctuations, that is, $\langle\cos(2Ql-\sqrt{2}\theta_a)\rangle=0$.
On the other hand, the recent DMRG simulation has rather confirmed
local patterns in the incommensurate vortex phase \cite{PhysRevLett.115.190402}.
Such a difference between the bosonization and DMRG may be related to
the fact that the DMRG simulation is performed at an open boundary condition while
the bosonization is done at the periodic boundary condition.

On the other hand, in  the commensurate flux case,
the situation is different. This is due to the fact that
in the presence of the umklapp terms, the axial U(1) symmetry is explicitly 
broken \cite{PhysRevA.92.013625}.
Below, we show that this explicit breaking of the continuous symmetry rather
supports the fixing of the relative phase.

We  first consider the case of $Q=\pi/2$
($\phi\approx \pi$).
In this case, due to the umklapp scattering, the following term
in the Hamiltonian is induced
at the mean-field level \cite{PhysRevA.92.013625}:
$g\cos2(\bar{\theta}_+-\bar{\theta}_{-})$ with a positive $g$.
Thus, $\bar{\theta}_+-\bar{\theta}_{-}=\frac{2m+1}{2}\pi$ in the ground state,
meaning that even at the mean-field level, we obtain
\beq
\langle n_{l,1} \rangle=\langle n_{l,2}\rangle=\frac{n}{2}.
\eeq
We point out that the above relation holds at
the bosonization level
since the umklapp scattering introduces $\cos2\sqrt{2}\theta_a$
which is always relevant in a weak coupling and locks $\theta_a$
\cite{PhysRevA.92.013625}.
This behavior around $\phi\approx\pi$ is consistent with the DMRG
simulation \cite{PhysRevLett.115.190402}.

We also look at the other commensurate cases where $Q=\frac{p}{q}\pi$
with two mutually prime integers $p,q$.
As discussed in Refs. \cite{PhysRevB.64.144515,tokuno},
by considering perturbation theory to order $q$,
the seemingly irrelevant term $\cos(\sqrt{2}\theta_a-2Ql)$
induces a term,
\beq
\cos(q\sqrt{2}\theta_a),
\eeq
which is relevant perturbation.
Thus, $\theta_a$ is fixed, which leads to a periodic pattern of
the local density.
An interesting point here is that the local periodic pattern
in density
is induced as a consequence of the breaking of the discrete symmetry

We also note that the similar argument is possible for the local current.
Namely, in the incommensurate vortex (or vortex fluid) phase, a local current pattern that
emerged in the mean-field theory disappears in the bosonization treatment.
In the commensurate vortex (or vortex lattice) phase, such a local current pattern is allowed
within the bosonization
due to the breaking of the discrete symmetry.

\subsection{Chiral-current reversal}
Recently, Greschner et al. revealed that by means of the DMRG,
the sign of the chiral current can be changed in the vicinity of the
commensurate vortex (or vortex lattice) phase \cite{PhysRevLett.115.190402}.
Here,
we interpret such a chiral-current reversal
in terms of our formalism
\footnote{The chiral-current reversal in the regime $J\gg K,U$
  has also been implicitly
  discussed in Ref. \cite{PhysRevB.64.144515}.}.

We first consider the case in the vicinity of $\phi=\pi$
($Q=\frac{\pi}{2}$).
When $\phi$ is identical to $\pi$, 
since the chiral current is given by
$j_c(l)=2nJ\sin\xi_{\pi/2}(-1)^l$,
the net chiral current
vanishes \cite{tokuno},
\beq
j_c\equiv\frac{1}{N}\sum_{l}j_c(l)=0.
\eeq
We now look at a regime where the flux value is given by $\phi=\pi+\delta$
and $\delta$ is small in such a way that the system is in the commensurate
vortex phase.
The intrachain hopping is then bosonized as
\beq
&&-J\sum_l\Big[e^{i(\frac{\pi+\delta}{2})}b_{l+1,1}^{\dagger}b_{l,1}
  +e^{-i(\frac{\pi+\delta}{2})}b_{l+1,2}^{\dagger}b_{l,2}\Big]
\nonumber\\
&&\sim-4nJ\sin^2\frac{\xi_{\pi/2}}{2}\cos\frac{\nabla\theta_s}{\sqrt{2}}
\cos\left(\frac{2\pi+\delta}{2}-\frac{\nabla\theta_a}{\sqrt{2}}\right)
\nonumber\\
&&-4nJ\cos^2\frac{\xi_{\pi/2}}{2}\cos\frac{\nabla\theta_s}{\sqrt{2}}
\cos\left(\frac{\delta}{2}+\frac{\nabla\theta_a}{\sqrt{2}}\right),
\eeq
where we neglect the terms containing $e^{\pm iQl}$,
since they do not contribute to the net chiral current.
Thus, the net chiral current is  given by
\beq
j_c=
J\left[\cos^2\frac{\xi_{\pi/2}}{2}
-\sin^2\frac{\xi_{\pi/2}}{2}\right]\delta +O(\delta^2),
\eeq
where we used the fact that $\theta_a$
is locked in the commensurate vortex phase.
Thus, it turns out that the chiral-current reversal occurs around
$\phi\approx\pi$ since $ j_c$ is negative for $\delta<0$,
which has the opposite sign compared to
the chiral current in the incommensurate vortex phase.

We next consider the other commensurate cases.
Then, it turns out that in such a case,
in addition to $\delta$ dependent terms, there is a
$\delta$ independent term
\cite{tokuno}.
Thus, to discuss the chiral current reversal
in the $Q\ne\frac{\pi}{2}$ case, we have to compare
the magnitudes of the different terms, each of which has a different sign.
Within the bosonization method, however, it is difficult to
obtain a solid answer in $Q\ne\frac{\pi}{2}$.

\subsection{Long-range interaction along rung direction:
synthetic dimension}
From the perspective of the synthetic dimension,
we introduce the interchain interaction, $V$.
As mentioned in Sec. II,
the coupling $V$ takes a  value similar to the onsite coupling $U$.
Therefore, we first look at the interaction terms in the limit $U=V$, which
is given by
\beq
\frac{U}{2}\sum_{l}[(n_{l,1}+n_{l,2})^2-(n_{l,1}+n_{l,2})].
\eeq
Apparently, the above term has SU(2) symmetry, and therefore
one may expect that the ferromagnetic ground state is
due to the spontaneous symmetry breaking of SU(2)
as in the case of the two-component bosons \cite{pethick2002bose}.
While the above argument may give some insight into what happens in the
presence of $V$, one must care about the fact that
the kinetic term in the Hamiltonian explicitly breaks
the SU(2) symmetry. 

To determine the phase structure,
we now use  the GP approach 
\footnote{See Ref. \cite{PhysRevA.92.053623} on discussions on more general
  long range interactions.}.
In the case of an incommensurate $Q$,
 the mean-field energy is obtained as
\beq
&&E_{-}(Q)+\frac{nU}{2}\left[\frac{1}{2}\sin^2\xi_Q\sin^22\gamma+1\right]\nonumber\\
&&+\frac{n\Delta}{2}\left[\left(-\frac{1}{4}\sin^2\xi_Q+\frac{1}{2}
  \right)\sin^22\gamma+\frac{\sin^2\xi_Q}{2}\right],
\label{eq:lri}
\eeq
where $E_-(Q)$ is the single particle spectrum in the lower band at $k=Q$,
$\Delta=V-U$ describes the mismatch from the isotropic limit.
Notice that as in the case of  $V=0$ \cite{PhysRevA.92.013625},
the energy does not
depend on the phase factor $\bar{\theta}_{\pm}$, which is a
direct consequence of
the absence of the umklapp scattering between $k=\pm Q$.
As far as a weak-coupling regime is concerned,
the number of minima in the lower band plays a crucial role in
determining the phase structure.
In the single minimum case, as in  $V=0$,
the Meissner phase is obtained.
To discuss the two minimum case,
we consider the regime where the mismatch is small
$|\Delta/U|\ll1$, considering the experiment with $^{87}$Rb \cite{stuhl}.
We then find through Eq. \eqref{eq:lri}
that in contrast to the $V=0$ case,
the biased ladder phase is favored
\footnote{The vortex phase is also allowed if 
  $\sin^2\xi_Q\ll 1$ and $\Delta<0$.}.
In accord with the analysis in Ref. \cite{PhysRevA.92.013625},
one can also discuss the quantum fluctuation effects
from the above mean-field analysis. Then, it is straightforward to show that
the phase structure does not change by the fluctuations
while a Bose-Einstein condensate predicted at the mean-field level
is substituted by the TLL
in both  the Meissner and biased-ladder phases.

We next consider the $Q=\frac{\pi}{2}$ case, which
is dominant commensurability in the weak-coupling limit.
By using the GP ansatz,
we obtain 
\beq
&&E_-(Q)+\frac{nU}{8}\Big[\frac{1+\cos2(\bar{\theta}_+-\bar{\theta}_-)}{2}
  \sin^2\xi_{Q}\sin^22\gamma+1\Big]\nonumber\\
&&+\frac{n\Delta}{2}\Bigg[\left(\frac{-1+\cos2(\bar{\theta}_+
    -\bar{\theta}_-)}{4}
  \sin^2\xi_Q+\frac{1}{2}
  \right)\sin^22\gamma\nonumber\\
&&  +\frac{1}{2}\sin^2\xi_Q\Bigg],
\eeq
where the relative phase dependence comes out
via the umklapp scattering between $k=\pm Q$.
In the ground state, $\cos2(\bar{\theta}_+-\bar{\theta}_-)=-1$
if  $|\Delta/U|$ is small. 
We then find that the vortex phase is stabilized if
$\Delta<0$, and the population-imbalance phase is stabilized if $\Delta\ge0$.
Notice also that the net chiral current at $\phi=\pi$
is zero regardless of the phases realized since it is proportional to
$\sin\left(\frac{\pi}{2}+Q\right)$ \cite{PhysRevA.92.013625}.
A difference between the vortex and population-imbalance phases
is that the vortex phase shows the local-current pattern while
the population-imbalance phase does not.

In total, through the analysis with the nonzero $V$,
it has been found that
the population-imbalance solution is widely favored.
In the presence of the umklapp scattering, which favors the population-balance
solution, the transition between the population imbalance and balance
solutions is also allowed as a function of $\Delta$.

\section{Summary}
In this paper, we considered the two-leg bosonic ladder model
in the presence of a magnetic flux especially
in the weak-coupling regime where a Bose-Einstein condensate is absent.
Commensurability on a flux
plays an important role in determining the low-energy properties,
which is in sharp contrast to
the Bose system with the spin-orbit coupling in a continuum space,
where the C-IC transition is absent \cite{lin2011spin}.
As long as the interaction strength is larger than the temperature,
the C-IC transition emerges even in a weak-coupling limit unlike
the Mott insulator in the Bose-Hubbard model.  
Our formalism may be useful to discuss the three (or more) leg ladder
case \cite{PhysRevB.91.054520,kolley,barbarino1,barbarino2}.

\section*{Acknowledgement}
The author is grateful to I. Danshita, T. Giamarchi,
M. Piraud, and A. Tokuno for discussions.


%

\end{document}